\documentclass[a4paper]{article}
\pdfoutput=1

\usepackage{graphicx,wrapfig}
\usepackage{lipsum}
\usepackage{amsmath,amssymb,mathrsfs}
\usepackage{hyperref}
\usepackage{url}
\usepackage{mathtools}
\usepackage{changepage}
\usepackage{multirow}
\usepackage{wrapfig}
\usepackage{subfig}
\usepackage{color}
\usepackage{epsfig,enumerate,amsmath,amsfonts,amssymb,amsthm,mathrsfs,ifpdf}
\usepackage{indentfirst,relsize}
\usepackage[numbers]{natbib}
\usepackage{setspace,graphicx}
\usepackage{latexsym}
\usepackage[all]{xy}
\usepackage[usenames,dvipsnames]{pstricks}
\usepackage{pst-grad} 
\usepackage{pst-plot} 
\usepackage[margin = 2.50cm]{geometry}

\usepackage{algorithm}

\usepackage{algpseudocode}

\newcommand{\remove}[1]{}

\newtheorem{theo}{Theorem}
\newtheorem{lem}[theo]{Lemma}

\newtheorem{coro}[theo]{Corollary}

\newtheorem{cl}[theo]{Claim}
\newcounter{Ca}[theo]
\newtheorem{ca}[Ca]{Case}
\newtheorem{defi}[theo]{Definition}
\newtheorem{remk}{Remark}

\title{Transitivity on subclasses of bipartite graphs}

\author{Subhabrata Paul \and Kamal Santra }

\author{Subhabrata Paul\footnote{Department of Mathematics, IIT Patna, India} \and Kamal Santra\footnote{Department of Mathematics, IIT Patna, India} }

\date{}

\begin{document}

\maketitle
\begin{abstract}
Let $G=(V, E)$ be a graph where $V$ and $E$ are the vertex and edge set, respectively. For two disjoint subsets $A$ and $B$, we say $A$ dominates $B$ if every vertex of $B$ is adjacent to at least one vertex of $A$. A vertex partition $\pi = \{V_1, V_2, \ldots, V_k\}$ of $G$ is called a \emph{transitive $k$-partition} if $V_i$ dominates $V_j$ for all $i,j$ where $1\leq i<j\leq k$. The maximum integer $k$ for which the above partition exists is called \emph{transitivity} of $G$ and it is denoted by $Tr(G)$. The \textsc{Maximum Transitivity Problem} is to find a transitive partition of a given graph with the maximum number of partitions. It was known that the decision version of \textsc{Maximum Transitivity Problem} is NP-complete for general graphs, which was proved by Hedetniemi et al. [Iterated colorings of graphs, \emph{Discrete Mathematics}, 278, 2004]. This paper first strengthens the NP-completeness result by showing that this problem remains NP-complete for perfect elimination bipartite graphs. On the other hand, we propose a linear-time algorithm for finding the transitivity of a given bipartite chain graph. We then characterize graphs with transitivity at least $t$ for any integer $t$. This result answers two open questions posed by J. T. Hedetniemi and S. T. Hedetniemi [The transitivity of a graph, \emph{J. Combin. Math. Combin. Comput}, 104, 2018].


\end{abstract}

\section{Introduction}
Let $G=(V, E)$ be a graph where $V$ and $E$ are the vertex and edge set, respectively. The \emph{neighbourhood} of a vertex $v\in V$ is the set of all adjacent vertices of $v$ and is denoted as $N(v)$. The \emph{degree} of a vertex, denoted as $d(v)$, is the number of edges incident to $v$. A vertex $v$ is said to \emph{dominates} itself and all its neighbouring vertices. A \emph{dominating set} of $G=(V,E)$ is a subset of vertices $D$ such that every vertex $x\in V\setminus D$ has a neighbour $y\in D$, i.e., $x$ is dominated by some vertex $y$ of $D$. For two disjoint subsets $A$ and $B$, we say $A$ \emph{dominates} $B$ if every vertex of $B$ is adjacent to at least one vertex of $A$. Over the past few decades, researchers have studied graph partitioning problem where the goal is to partition the vertex set (or edge set) into some parts with desired properties, such as independence, having minimum edges across partite sets etc. In this paper, we have studied a special type of graph partitioning problem. We are interested in partitioning the vertex set into some parts such that the partite sets follow different types of domination properties.

Cockayne and Hedetniemi, in 1977, introduced the notion of \emph{domatic partition} of a graph $G=(V,E)$ where the vertex set is partitioned into $k$ parts, say $\pi =\{V_1,V_2, \ldots, V_k\}$, such that each $V_i$ is a dominating set of $G$ \cite{cockayne1977towards}. The maximum order of such a domatic partition is called \emph{domatic number} of $G$ and it is denoted by $d(G)$. Another similar type of partitioning problem is \emph{Grundy coloring}. Christen and Selkow introduced Grundy coloring of a graph $G=(V,E)$ in 1979 \cite{CHRISTEN197949}. In Grundy coloring problem, the vertex set is partitioned into $k$ parts, say $\pi =\{V_1,V_2, \ldots, V_k\}$, such that each $V_i$ is an independent set and for all $1\leq i< j\leq k$, $V_i$ dominates $V_j$. The maximum order of such a domatic partition is called \emph{Grundy number} of $G$ and it is denoted by $\Gamma(G)$. In 2004, S. M. Hedetniemi et al. introduced another such partitioning problem, namely \emph{upper iterated domination partition}\cite{erdos2003equality}. In an upper iterated domination partition, the vertex set is partitioned into $k$ parts, say $\pi =\{V_1,V_2, \ldots, V_k\}$, such that for each $1\leq i\leq k$, $V_i$ is a minimal dominating set of $G\setminus (\cup_{j=1}^{i-1} V_j)$. The \emph{upper iterated domination number}, denoted by $\Gamma^*(G)$, is equals to the maximum order of such a vertex partition. Recently, in 2018, Haynes et al. generalized the idea of \emph{domatic partition} and introduced the concept of \emph{upper domatic partition} of a graph $G$ where the vertex set is partitioned into $k$ parts, say $\pi =\{V_1,V_2, \ldots, V_k\}$, such that for each $i, j$, $1\leq i<j\leq k$, either $V_i$ dominates $V_j$ or $V_j$ dominates $V_i$ or both \cite{haynes2020upper}. The maximum order of such a upper domatic partition is called \emph{upper domatic number} of $G$ and it is denoted by $D(G)$. All these problem, domatic number \cite{chang1994domatic,zelinka1980domatically,zelinka1983k}, Grundy number \cite{zaker2005grundy,zaker2006results,furedi2008inequalities,hedetniemi1982linear, effantin2017note}, upper iterated number \cite{erdos2003equality}, upper domatic number \cite{haynes2020upper} have been extensively studied both from algorithmic and structural point of view.

In this article, we have studied a similar graph partitioning problem, namely \emph{transitive partition}. In 2018, J. T. Hedetniemi and S. T. Hedetniemi \cite{hedetniemi2018transitivity} introduced this notion as a generalization of Grundy coloring. A \emph{transitive $k$-partition} is defined as a partition of the vertex set into $k$ parts, say $\pi =\{V_1,V_2, \ldots, V_k\}$, such that for all $1\leq i< j\leq k$, $V_i$ dominates $V_j$. The maximum order of such a transitive partition is called \emph{transitivity} of $G$ and is denoted by $Tr(G)$. The \textsc{Maximum Transitivity Problem} and its correcponding decision version is defined as follows:\\

\noindent\textsc{\underline{Maximum Transitivity Problem(MTP)}}

\noindent\emph{Instance:} A graph $G=(V,E)$

\noindent\emph{Solution:} A transitive partition of $G$

\noindent\emph{Measure:} Order of the transitive partition of $G$\\

\noindent\textsc{\underline{Maximum Transitivity Decision Problem(MTDP)}}

\noindent\emph{Instance:} A graph $G=(V,E)$, integer $k$

\noindent\emph{Question:} Does $G$ have a transitive partition of order at least $k$?\\

\noindent Note that a Grundy coloring is a transitive partition with addition restriction that each partite set must be independent. In a domatic partition $\pi =\{V_1,V_2, \ldots, V_k\}$ of $G$, since each partite sets are dominating sets of $G$, we have domination property in both directions, i.e., $V_i$ dominates $V_j$ and $V_j$ dominates $V_i$ for $1\leq i< j\leq k$. Whereas in a transitive partition $\pi =\{V_1,V_2, \ldots, V_k\}$ of $G$, we have domination property in one direction, i.e., $V_i$ dominates $V_j$ for $1\leq i< j\leq k$. In a upper domatic partition $\pi =\{V_1,V_2, \ldots, V_k\}$ of $G$, for $1\leq i<j\leq k$, either $V_i$ dominates $V_j$ or $V_j$ dominates $V_i$ or both. The definition of each vertex partitioning problem ensures the following inequality for any graph $G$. For any graph $G$, $1\leq d(G)\leq \Gamma(G)\leq \Gamma^*(G)\leq Tr(G)\leq D(G)\leq n$.

In the introductory paper, J. T. Hedetniemi and S. T. Hedetniemi \cite{hedetniemi2018transitivity} showed that the transitivity of a graph $G$ bounded by $\Delta+1$, where $\Delta$ maximum degree of $G$. They also proved a necessary and sufficient condition for graphs with $Tr(G)=2$ and graphs with $Tr(G)\geq 3$. They further showed that transitivity and Grundy number are the same for trees. Therefore, the linear time algorithm for finding the Grundy number of a tree, presented in \cite{hedetniemi1982linear}, implies that we can find the transitivity of a tree in linear time as well. Moreover, for any graph, transitivity is equal to upper iterated domination number \cite{hedetniemi2018transitivity}, and the decision version of the upper iterated domination problem is known to be NP-complete \cite{hedetniemi2004iterated}. Therefore, MTDP is NP-complete for chordal graphs. It is also known that every connected graph $G$ with $Tr(G)\geq 4$ has a transitive partition $\pi =\{V_1,V_2, \ldots, V_k\}$ such that $\lvert V_k \rvert$=$\lvert V_{k-1} \rvert=1$ and $\lvert V_{k-i} \rvert \leq 2^{i-1}$ for $2\leq i\leq k-2$ \cite{haynes2017transitivity}. This implies that MTP is fixed-parameter tractable.

In this paper, we study the computational complexity of the transitivity problem and also prove a few structural results. The main contributions are summarized below:

\begin{enumerate}
	\item[1.] We show that MTDP is NP-complete for bipartite graphs. As the resultant graph in the polynomial reduction is a perfect elimination bipartite graph, MTDP remains NP-complete for this important subclass of bipartite graphs.
	
	\item[2.] We prove that finding transitivity of a given bipartite chain graph $G$ is the same as finding the maximum index $t$ such that $G$ contains either $K_{t,t}$ or  $K_{t,t}-\{e\}$ as an induced subgraph. We design a linear time algorithm for MTP in a bipartite chain graph based on this fact.
	
	\item[3.] In \cite{hedetniemi2018transitivity}, the authors posed two open problems of characterizing graphs with $Tr(G)\geq 4$ and graphs with $Tr(G)=3$. We solve these open problems by giving a general characterization of graphs with $Tr(G)\geq t$, for any integer $t$.
\end{enumerate}

The rest of the paper is organized as follows. In Section 2, we present the NP-completeness of MTDP for perfect elimination bipartite graphs. Then in Section 3, we design a linear time algorithm for MTP in bipartite chain graphs. Section 4 deals with characterization of graphs with $Tr(G)$ for any integer $t$. Finally, Section 5 concludes the paper.

\section{NP-complete for bipartite graphs}

In this section, we show that \textsc{Maximum Transitivity Decision Problem} is NP-complete for bipartite graphs. Clearly, this problem is in NP. We prove the NP-completeness of this problem by showing a polynomial time reduction from  \textsc{Proper $3$-Coloring Decision Problem} in graphs. A proper $k$-colring of a graph $G=(V,E)$ is a funtion  $g$, from $V$ to $\{1,2,3, \ldots , k\}$ such that $g(u)\not= g(v) $ for any edge $uv \in E$. The \textsc{Proper $3$-Coloring Decision Problem} is defined as follows:

\noindent\textsc{\underline{Proper $3$-Coloring Decision Problem}}

\noindent\emph{Instance:} A graph $G=(V,E)$

\noindent\emph{Question:} Does there exist a proper $3$-coloring of $V$?

The \textsc{Proper $3$-Coloring Decision Problem} is known to be NP-complete \cite{garey1990guide}. Given an instance of \textsc{Proper $3$-Coloring Decision Problem}, say $G=(V,E)$, we construct an instance of MTDP. The construction is as follows:

\noindent\textbf{Construction:}
Let $V=\{v_1, v_2, \ldots, v_n\}$ and $E= \{e_1, e_2, \ldots, e_m\}$.
\begin{itemize}
	\item[$1$]  For each vertex $v_i\in V$, we consider two paths of length three, $P_{v_i}=\{x_i,w_i,v_i,z_i\}$ and $P_{v_i}^\prime=\{x_i^\prime,w_i^\prime,v_i^\prime,z_i^\prime\}$ in $G'$, where $x_i$ and $z_i$ are the pendant vertices of $P_{v_i}$ and  $x'_i$ and $z'_i$ are the pendant vertices of $P'_{v_i}$. Similarly, for each edge $e_j\in E$, we consider two paths of length three,  $P_{e_j}=\{x_{e_j},w_{e_j},v_{e_j},z_{e_j}\}$ and $P_{e_j}^\prime=\{x_{e_j}^\prime,w_{e_j}^\prime,v_{e_j}^\prime,z_{e_j}^\prime\}$ in $G'$. Next consider six more paths of length three, $P_a=\{x_{a},w_{a},v_{a},z_{a}\}$,  $P'_a=\{x_{a}^\prime,w_{a}^\prime,v_{a}^\prime,z_{a}^\prime\}$, $P_b=\{x_{b},w_{b},v_{b},z_{b}\}$,  $P'_b=\{x_{b}^\prime,w_{b}^\prime,v_{b}^\prime,z_{b}^\prime\}$, $P_e=\{x_{e},w_{e},v_{e},z_{e}\}$ and $P'_e=\{x_{e}^\prime,w_{e}^\prime,v_{e}^\prime,z_{e}^\prime\}$ in $G'$.

	\item[$2$] For each edge $e_j\in E$, we take two vertices, $e_j$  and $e'_j$ in $G'$ and also we take two extra vertices $e$ and $e'$ in $G'$. Let  $A=\{e_1,e_2,\ldots ,e_m,e\}$ and  $B=\{e_1^\prime,e_2^\prime,\ldots,e_m^\prime,e^\prime\}$. We make a complete bipartite graph with vertex set $A\cup B$. 
	
	\item[$3$] Next we add the following edges: for every edge $e_k=(v_i,v_j)\in E$, we join the edges $(v_i,e_k)$, $(v_j, e_k)$, $(v_{e_k}, e_k)$, $(v'_i,e'_k)$, $(v'_j, e'_k)$ and $(v'_{e_k},e'_k)$. Also we add the edges $(v_a,e)$, $(v_b, e)$, $(v_{e}, e)$, $(v'_a,e')$, $(v'_b, e')$ and $(v'_{e},e')$.
	
	\item[$4$] Finally, we set $k=m+5$.
	
\end{itemize}

\begin{figure}[htbp!]
	\centering
	\includegraphics[scale=0.75]{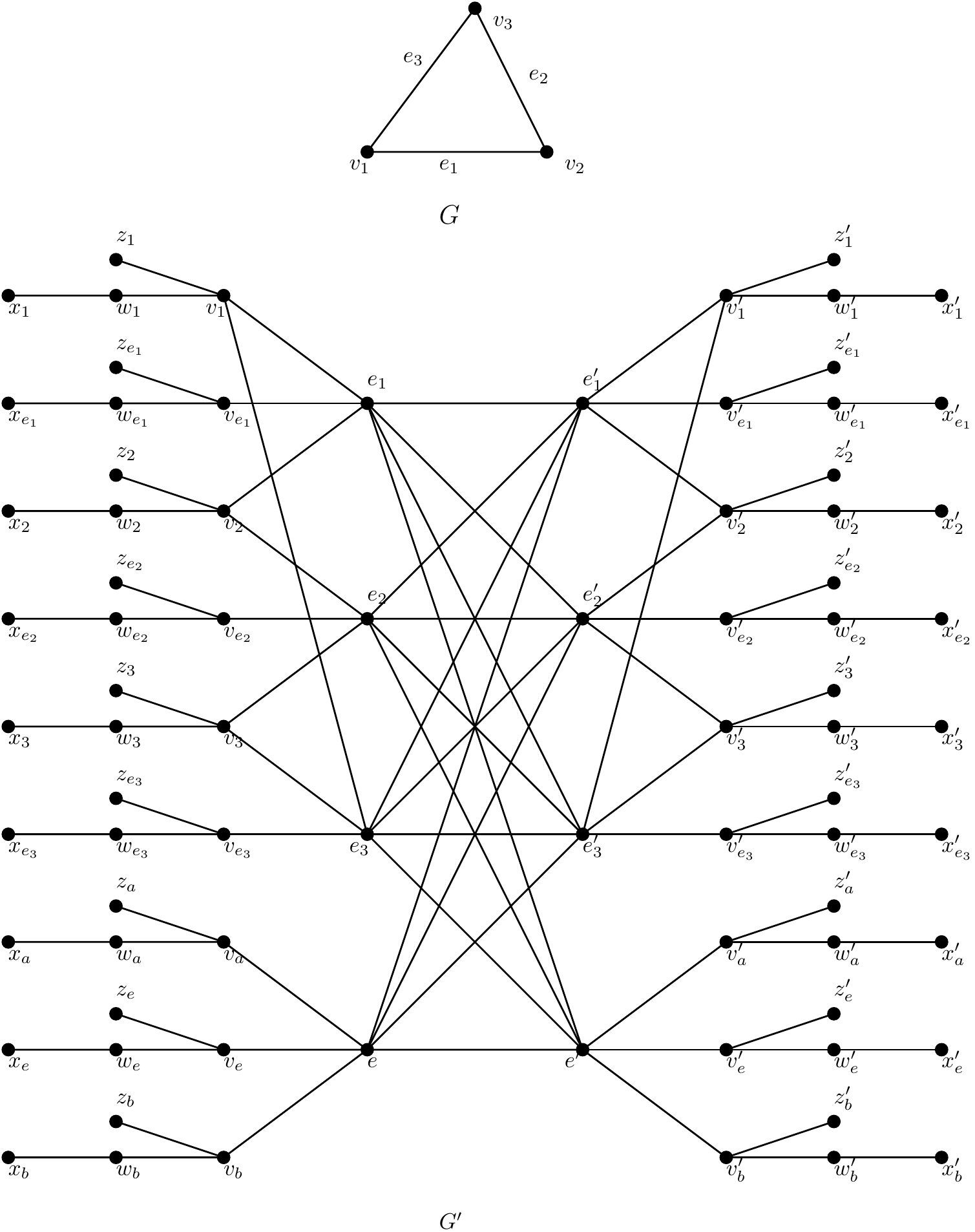}
	\caption{Construction of $G^\prime$}
	\label{fig:bipartitenp}
	
\end{figure}

From the above construction, it is clear that the graph $G'=(V', E')$ consists of $10m+8n+26$ vertices and $m^2+14m+6n+25$ edges. The construction is illustrated in Figure \ref{fig:bipartitenp}.

Now we show that  $G$ has a proper 3-coloring if and only if $G^\prime$ has a transitive $k$-partition. For the forward direction, we have the following lemma.

\begin{lem}
	If $G=(V,E)$ has a proper 3-coloring, then $G'=(V', E')$ has a transitive $k$-partition.
\end{lem}
\begin{proof}
	Given a proper 3-coloring $g$ from $V$ to $\{1,2,3\}$, a partition $\pi=\{V_1,V_2, \ldots,V_{m+5}\}$ of the vertices of $G'$ can be obtained in the following ways:
	\begin{itemize}
		\item If $g(v_i)=q$ in $G$, then $v_i, v_i^\prime \in V_q$ and we put $v_a, v_a^\prime\in V_1$, and $v_b, v_b^\prime\in V_2$.
		\item For an edge $e_k=(v_i,v_j)$ in $G$, we put $v_{e_k}, v_{e_k}^\prime \in V_l$, where $l= \{1, 2, 3\} \setminus \{g(v_i),g(v_j)\}$. Also, we put $v_e, v_e^\prime\in V_3$. The other vertices of all $P_4$ and $P_4^\prime$, i.e. $\{x,w,z\}$ and $\{x',w',z'\}$, are put to a partition based on the partition of $v$ or $v'$, respectively as shown in Figure \ref{fig:vertexedge_coloring}.
		\begin{figure}[htbp!]
			\centering
			\includegraphics[scale=0.70]{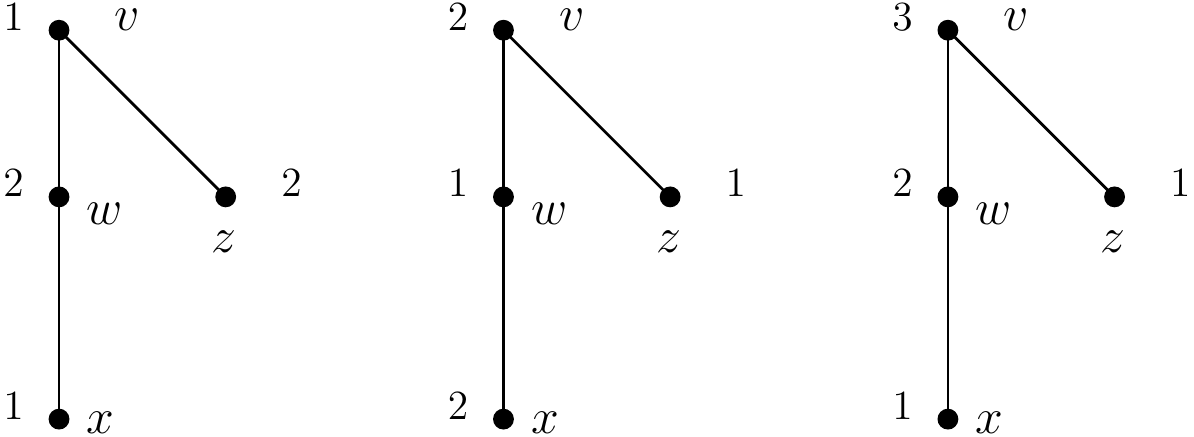}
			\caption{Partition of all $P_4$}
			\label{fig:vertexedge_coloring}
		\end{figure}
		\item Lastly, we put $e_j, e_j^\prime \in V_{3+j}$, for all $1\leq j\leq m$, and $e\in V_{m+4}$, $e^\prime \in V_{m+5}$.
	\end{itemize}

	Now we show that the above process produces a transitive $k$-partition $\pi=\{V_1,V_2, \ldots,V_{m+5}\}$ of $G'$. Let $H$ be the complete bipartite graph induced by $A\cup B$. Since $H$ is a complete bipartite graph, then $V_i$ dominates $V_j$ for $4\leq i<j\leq m+5$. Also each vertex from $A\cup B$ is adjacent to a vertex from each $V_1, V_2$ and $V_3$. Hence $V_i$ dominates $V_t$, for all $i=1,2,3$ and $t>3$. Also from Figure \ref{fig:vertexedge_coloring}, it is clear that $V_i$ dominates $V_j$ for $1\leq i<j\leq 3$. Therefore, $\pi$ is a transitive $k$-partition of $G'$. 
\end{proof}

The following lemma shows that the converse of the statement is also true.

\begin{lem}
	If $G^\prime$ has a transitive $k$-partition, then $G$ has a proper 3-coloring.
\end{lem}

\begin{proof}
	Let $\pi=\{V_1,V_2,\ldots ,V_k\}$ be a transitive $k$-partition of $G'$. By Proposition 11 of \cite{hedetniemi2018transitivity}, we know that $\pi$ can be transformed into $\pi'=\{V'_1,V'_2,\ldots ,V'_k\}$, such that $\lvert V'_k \rvert=\lvert V'_{k-1}\rvert =1$. So, without loss of generality, let us assume $G$ has a transitive $k$-partition $\pi=\{V_1,V_2,\ldots ,V_k\}$, such that $\lvert V_k \rvert =\lvert V_{k-1} \rvert=1$.\\

	\begin{cl}\label{cl:PartitionOfV'}
		In the transitive $k$-partition $\pi$, the partitions $\{V_1, V_2, V_3\}$ contain only vertices from $V'\setminus (A\cup B)$ and the partitions $\{V_4, V_5,\ldots V_k\}$ contain only vertices from $A\cup B$.
	\end{cl}
	
	\begin{proof}
		We divide the proof into the following four cases based on the partition of $e$ and $e'$:
		
		\begin{ca}\label{cas}
			\textbf{$e\in V_{m+5}$ and $e'\in V_{m+4}$}
			
			\textnormal{
				Note that $\{v_a,v_b, v_e\}$ cannot be in $V_p$ for $p\geq 4$ because in that case $e$ would be in $V_3$. Similarly, $\{v'_a,v'_b,v'_e\}$ cannot be in $V_p$ for $p\geq 4$ as well. Therefore, the  vertices from $\{v_a,v_b,v_e\}$ and $\{v'_a,v'_b,v'_e\}$ belong to $V_p$ for $1\leq p \leq 3$. To dominate $e$ and $e'$, the vertices from $\{e'_1, e'_2, \ldots, e'_m\}$  and  $\{e_1, e_2, \ldots, e_m\}$ must belong to $\{V_4, V_5, \ldots, V_{m+3}\}$, respectively. Moreover, for $4\leq i \leq m+3$, each $V_i$ contains exactly one vertex from $\{e'_1, e'_2, \ldots, e'_m\} $ to dominate $e$ and exactly one vertex from $\{e_1, e_2, \ldots, e_m\}$ to dominate $e'$. Hence, the vertices of $A\cup B$ belong to $\{V_4, V_5,\ldots V_{m+5}\}$. Note that none of the vertices from $\{v_1, v_2, \ldots, v_n, v'_1, v'_2, \ldots, v'_n, v_{e_1}, v_{e_2}, \ldots, v_{e_m}, v'_{e_1}, v'_{e_2}, \ldots, v'_{e_m}\}$ can belong to $V_p$ for $p\geq 4$ because otherwise there exists a vertex in $A\cup B$ which belongs to $V_3$. Since degree of every other vertices is at most $2$, they cannot belong to $V_p$ for $p\geq 4$. Hence, $\{V_4, V_5,\ldots V_k\}$ contain only vertices from $A\cup B$ and $\{V_1, V_2, V_3\}$ contain only vertices from $V'\setminus (A\cup B)$.
			}
		\end{ca}

		\begin{ca}
			\textbf{$e\in V_{m+5}$ and $e'\notin V_{m+4}$}
			
			\textnormal{
				Using similar arguments as in the previous case, we know that vertices from $\{v_a,v_b,v_e\}$ belong to $V_p$ for $1 \leq p \leq  3$ and the vertices from $\{e'_1, e'_2, \ldots, e'_m,e'\} $ belong to $\{V_4, V_5, \ldots, V_{m+3},V_{m+4}\}$ to dominate $e$. Moreover, every $V_i$ for $4\leq i \leq m+4$ contains exatly one vertex from $\{e'_1, e'_2, \ldots, e'_m,e'\} $. Since the vertices from $\{e'_1, e'_2, \ldots, e'_m,e'\} $ belong to $\{V_4, V_5, \ldots, V_{m+3},V_{m+4}\}$, no vertex from $\{v'_1, v'_2, \ldots, v'_n, v'_{e_1}, v'_{e_2}, \ldots, v'_{e_m},\\ v'_a, v'_b, v'_e\}$ can be in $V_p$ for $p\geq 4$. As $e'\notin V_{m+4}$, there exist a vertex from $\{e'_1, e'_2, \ldots, e'_m\} $, say $e'_j$ such that $e'_j \in V_{m+4}$. To dominate $e'_j$, the vertices from  $\{e_1, e_2, \ldots, e_m\}$  belong to $\{V_4, V_5, \ldots, V_{m+3}\}$. With similar arguments as in previous case, we know that every vertex of $\{v_1, v_2, \ldots, v_n, v_{e_1}, v_{e_2}, \ldots, v_{e_m}\}$ belong to $V_p$ for $1\leq p\leq 3$. Hence, $\{V_4, V_5,\ldots V_k\}$ contain only vertices from $A\cup B$ and $\{V_1, V_2, V_3\}$ contain only vertices from $V'\setminus (A\cup B)$.
			}
		\end{ca}

		\begin{ca}
			\textbf{$e\notin V_{m+5}$ and $e'\in V_{m+4}$}
			
			\textnormal{		
				Using similar arguments as in Case \ref{cas}, we know that the vertices from $\{v'_a,v'_b,v'_e\}$ belong to $V_p$ for $1 \leq p \leq  3$ and the vertices from $\{e_1, e_2, \ldots, e_m,e\} $ belong to $\{V_4, V_5, \ldots, V_{m+3}\}$. As $e\notin V_{m+5}$, there exist a vertex from $\{e_1, e_2, \ldots, e_m\} $, say $e_j$ such that $e_j \in V_{m+5}$. Therefore, the vertices from $\{e_1, e_2, \ldots, e_m,e\} $ belong to $\{V_4, V_5, \ldots, V_{m+3},V_{m+5}\}$. Moreover, every $V_i$ for $4\leq i \leq m+5$ with $i\not=m+4$, contains exatly one vertex from $\{e_1, e_2, \ldots, e_m,e\}$. Since the vertices $\{e_1, e_2, \ldots, e_m,e\} $ belong to $\{V_4, V_5, \ldots, V_{m+3},V_{m+5}\}$, no vertex from $\{v_1, v_2, \ldots, v_n, v_{e_1}, v_{e_2}, \ldots, v_{e_m}, v_a, v_b, v_e\}$ can be in $V_p$ for $p\geq 4$. Since $e_j \in V_{m+5}$, the vertices from  $\{e'_1, e'_2, \ldots, e'_m\}$ must belong to $\{V_4, V_5, \ldots, V_{m+3}\}$ to dominate $e_j$. With similar arguments as in Case \ref{cas}, we know that every vertex of $\{v'_1, v'_2, \ldots, v'_n, v'_{e_1}, v'_{e_2}, \ldots, v'_{e_m}\}$ belong to $V_p$ for $1\leq p\leq 3$. Hence, $\{V_4, V_5,\ldots V_k\}$ contain only vertices from $A\cup B$ and $\{V_1, V_2, V_3\}$ contain only vertices from $V'\setminus (A\cup B)$.
			}
		\end{ca}		
		
		\begin{ca}
			\textbf{$e\notin V_{m+5}$ and $e'\notin V_{m+4}$}
			
			\textnormal{
				Since the degree of each vertex of $V'\setminus (A\cup B)$ is at most $m+2$, they cannot belong to $V_{m+4}$ or $V_{m+5}$, i.e., only verties from $A\cup B$ can be in $V_{m+4}$ or $V_{m+5}$. Also since we can interchange the vertices between $V_{m+4}$ and $V_{m+5}$ in the transitive partition, without loss of generality, we can assume that $e_1\in V_{m+5}$ and $e'_s \in V_{m+4}$ in the transitive $k$-partition of $G'$. Also, let $e_1$ be the edge between $v_1$ and $v_2$ in $G$ and in the transitive $k$-partition of $G'$, $v_1\in V_l$ and $v_2\in V_t$ where $l\geq t$.
				First we show that every vertex of the form $v'_r$ belong to a partition $V_p$ such that $p\leq l$. Because otherwise  if $v'_r \in V_{l+i}$ For some $i\geq 1$, then each of $V_3, V_4, \ldots, V_{l+i-1}$ contains at least one vertex from $B \setminus \{e'_s\}$.
				Also, to dominate $e_1$, each of $\{V_{l+1}, \ldots, V_{m+3}\}$ contains at least one vertex from $B \setminus \{e'_s\}$. This implies that each of the $(m+1)$ partitions, $\{V_3,V_4, \ldots, V_{m+3}\}$, contains one vertex from the set of $m$ vertices, $B \setminus \{e'_s\}$, which is impossible.
				Next we show that the vertex $e$ belong to a partition $V_t$ such that $t\geq l+1$. This is because to dominate $e'_s$, each of $\{V_{l+1}, \ldots, V_{m+3}\}$ contains at least one vertex from $A\setminus \{e_1\}$. Also, since $v_1\in V_l$, then each of $\{V_3, V_4, \ldots, V_{l-1}\}$ contains at least one vertex from $A\setminus \{e_1,e\}$. Therefore, $e$ belong to a partition $V_t$ such that $t\geq l+1$.
				Next we show that $l \leq 3$. Since degree of $e_1$ is $m+4$, each of its neighbours should belong to exactly of the the partitions $\{V_1, V_2, \ldots V_{m+4}\}$. Moreover, since $v_1\in V_l$, no vertex from $B$ can be in $V_l$. Also, no vertex of $\{v_a,v_b,v_e\}$ belongs to $V_l$ as $l\geq 4$. Hence, none of the neighbours of $e$ belong to $V_l$. But $e$ belongs to a partition $V_t$ where $t\geq l+1$. Therefore, $V_l$ cannot dominate $V_t$, which is a contradiction. 
				Now, $l\leq 3$ implies that the vertices of $B$ belong to $\{V_4, V_5, \ldots, V_{m+4}\}$ and Moreover, each of these partitions contains exatly one vertex from $B$. And since the vertices of $B$ belong to $\{V_4, V_5, \ldots, V_{m+4}\}$,  the vertices of $\{v'_1, v'_2, \ldots, v'_n, v'_{e_1}, v'_{e_2}, \ldots, v'_{e_m}, v'_a, v'_b, v'_e\}$ belongs to some partition $V_p$ for some $p\leq 3$. As $e'_s\in V_{m+4}$, to dominate $e'_s$, the vertices from $A\setminus \{e_1\}$ belong to $\{V_4, V_5, \ldots, V_{m+3}\}$. With similar arguments as in Case \ref{cas}, we know that each vertex of $\{v_1, v_2, \ldots, v_n, v_{e_1}, v_{e_2}, \ldots, v_{e_m}, v_a,v_b,v_e\}$ belong to $V_p$ for some $p\leq 3$. Hence, $\{V_4, V_5,\ldots V_k\}$ contain only vertices from $A\cup B$ and $\{V_1, V_2, V_3\}$ contain only vertices from $V'\setminus (A\cup B)$.
			}
		\end{ca}

		Therefore, for all the cases, the partitions $\{V_1, V_2, V_3\}$ contain only vertices from $V'\setminus (A\cup B)$ and the partitions $\{V_4, V_5,\ldots V_k\}$ contain only vertices from $A\cup B$.
	\end{proof}

	%

	Now let us define the mapping $g$ from $V$ to $\{1,2,3\}$. We set $g(v_i)=p$, if $v_i$ is in the partition $V_p$ in $G'$. By the previous claim, the mapping is well-defined. Let $e_k=(v_i,v_j)$ be an edge in $G$. By Claim \ref{cl:PartitionOfV'}, in $G'$ the vertex $e_k$ belongs to some partition $V_t$ where $t\in \{4,5,\ldots,k\}$. Moreover, since none of the vertices of $A\cup B$ belongs to $V_1, V_2$ or $V_3$, the vertices $v_i, v_j$ and $v_{e_k}$ must belong to three different partitions among $V_1, V_2, V_3$. Hence, $v_i$ and $v_j$ belong to different partitions among $V_1, V_2, V_3$. This imples that $g(v_i)\neq g(v_j)$. Therefore $g$ defines a proper $3$-coloring in $G$.
\end{proof}

Hence, we have the following theorem.

\begin{theo}
	The \textsc{Maximum Transitivity Decision Problem} is NP-complete for bipartite graphs.
\end{theo}

An edge $uv$ in a bipartite graph $G$ is called \emph{bisimplicial} if $N(u)\cup N(v)$ induces a biclique in $G$. For an edge ordering $(e_1,e_2, \ldots, e_k)$, let $S_i$ be the set of endpoints of $\{e_1,e_2, \ldots, e_i\}$ and $S_0=\emptyset$. An ordering $(e_1,e_2, \ldots, e_k)$ is a perfect edge elimination ordering for a bipartite graph $G=(V, E)$ if $G[V\setminus S_i]$ has no edges and each edge $e_i$ is a bisimplicial edge in $G[V\setminus S_i]$. A graph $G$ is a perfect elimination bipartite if and only if it admits a perfect edge elimination ordering \cite{golumbic1978perfect}. Note that, the construction implies that $G'$ in a perfect elimination bipartite graph. If we consider all the pendant edges of $G'$ in any order and after that a matching of the complete subgraph $H$, then these edges form a perfect edge elimination ordering of $G'$. Therefore, we have the following corollary. 

\begin{coro}\label{bipartite_coro1}
	The \textsc{Maximum Transitivity Decision Problem} remains NP-complete for perfect elimination bipartite graphs.
\end{coro}


	\section{Transitivity in bipartite chain graph}
		In this section, we design a linear time algorithm to solve the transitivity of a given bipartite chain graph. A bipartite graph $G=(X\cup Y,E)$ is called a \textit{bipartite chain graph} if there exist an ordering of vertices of $X$ and $Y$, say $\sigma_X= (x_1,x_2, \ldots ,x_m)$ and $\sigma_Y=(y_1,y_2, \ldots ,y_n)$, such that $N(x_m)\subseteq N(x_{m-1})\subseteq \ldots \subseteq N(x_2)\subseteq N(x_1)$ and $N(y_n)\subseteq N(y_{n-1})\subseteq \ldots \subseteq N(y_2)\subseteq N(y_1)$. This ordering of $X$ and $Y$ is called a chain ordering. A chain ordering of a bipartite chain graph can be computed in linear time \cite{heggernes2007linear}. To design the algorithm, we first prove that if $t$ is the maximum integer such that $G$ contains either $K_{t,t}$ or  $K_{t,t}-\{e\}$ as an induced subgraph, then $Tr(G)=t+1$. After that, we design an algorithm for finding maximum integer $t$ such that $G$ contains either $K_{t,t}$ or  $K_{t,t}-\{e\}$ as an induced subgraph.

	\begin{lem}\label{chain_th1}
		Let $G=(X\cup Y,E)$ be a chain graph and $t$ be the maximum integer such that $G$ contains either $K_{t,t}$ or  $K_{t,t}-\{e\}$ as an induced subgraph, then $Tr(G)=t+1$. 
	\end{lem}
	
	\begin{proof} 
		Suppose $t$ is the maximum integer such that $G=(X\cup Y,E)$ contains either $K_{t,t}$ or  $K_{t,t}-\{e\}$ as an induced subgraph. In this case, $Tr(G)\geq t+1$ because transitivity of $K_{t,t}$ or $K_{t,t}-\{e\}$ is $t+1$. Next, we show that $Tr(G)$ cannot be greater than $t+1$ by proving the following claim.
		
		\begin{cl}\label{chain_claim1}
			If $Tr(G)=m+1$, then $G=(X\cup Y,E)$ contains either $K_{m,m}$ or  $K_{m,m}-\{e\}$ as an induced subgraph.
		\end{cl}
		
		\emph{Proof of Claim \ref{chain_claim1}}. To prove this claim, we use induction on $m$. For $m=1$, i.e., $Tr(G)=2$, by the Proposition 5 of \cite{hedetniemi2018transitivity}, $G$ contains at least one edge. This implies that $G$ contains $K_{1,1}$ as an induced subgraph. Also, for $m=2$,  i.e., $Tr(G)=3$, $G$ contains either an induced $C_3$, an induced $P_4$ or an induced $C_4$ by the Proposition 7 of \cite{hedetniemi2018transitivity}. Since $G$ is a bipartite chain graph, it cannot contain $C_3$. Therefore $G$ contains either $P_4$, i.e., $K_{2,2}-\{e\}$ or $C_4$, i.e.,  $K_{2,2}$ as an induced subgraph. 
		
		By induction hypothesis let us assume that the claim is true for any graph $G$ with $Tr(G)<m+1$. Let $G$ be a bipartite chain graph with $Tr(G)=m+1$. Also let $\{V_1,V_2, \ldots ,V_{m+1}\}$ be a transitive $(m+1)$-partition of $G$. Let $G'=G[V_2\cup V_3\cup \ldots \cup V_{m+1} ]$ and so $Tr(G^\prime)=m$. By induction hypothesis $G'$ contains either $K_{m-1,m-1} $ or $K_{m-1,m-1}-\{e\}$ as an induced subgraph. Let $X'=\{x_1, x_2, \ldots , x_{m-1}\}$ and $Y'=\{y_1, y_2, \ldots , y_{m-1}\}$ be two sets such that $G'[X'\cup Y']$ is  either $K_{m-1,m-1} $ or $K_{m-1,m-1}-\{e\}$, where $e=x_{i}y_{j}$ for some $i$ and $j$ in $\{1,2, \ldots, m-1\}$. Now,  in $G$, since $V_1$ dominates $V_j$, for all  $j\geq 2$, $V_1$ contains at least one vertex from $X$ and also at least one vertex from $Y$. Let $\{x_{l_1},x_{l_2}, \ldots, x_{l_s}\}$ and $\{y_{k_1},y_{k_2}, \ldots, y_{k_t}\}$ be the vertices in $V_1$ from $X$ and $Y$ respectively. Since $G$ is a bipartite chain graph, there exist $x_p\in\{x_{l_1},x_{l_2}, \ldots, x_{l_s}\}$ and $ y_q\in \{y_{k_1},y_{k_2},\ldots, y_{k_t}\}$ such that $N(x_p)\supseteq N(x)$ for all $x\in\{x_{l_1}, \ldots, x_{l_s}\}$ and $N(y_q)\supseteq N(y)$ for all  $y \in\{y_{k_1},y_{k_2}, \ldots, y_{k_t}\}$.  Therefore, $x_p$ dominates $\{y_1,y_2, \ldots, y_{m-1}\}$ and $y_q$ dominates $\{x_1, x_2, \ldots, x_{m-1}\}$. Now, if $G'[X'\cup Y']$ is $K_{m-1,m-1} $,  then clearly $\{x_1,x_2, \ldots, x_{m-1},x_p\}$ and $\{y_1,y_2, \ldots, y_{m-1},y_q\}$ induces either a $K_{m,m} $ or a $K_{m,m}-\{e\}$ in $G$ depending on whether the edge $x_p y_q$ is in $G$ or not. On the other hand, if $G'[X'\cup Y']$ is $K_{m-1,m-1}-\{e\}$, where $e=x_{i}y_{j}$ for some $i$ and $j$ in $\{1,2, \ldots, m-1\}$. In this case also, we can argue in a similar way that there exist $x_p\in V_1$ and $y_q\in V_1$, such that $x_p$ dominates  $\{y_1,y_2, \ldots, y_{m-1}\}$ and $y_q$ dominates $\{x_1, x_2, \ldots, x_{m-1}\}$. Since $G$ is a bipartite chain graph, either $N(x_p)\subseteq N(x_i)$ or $N(x_p)\supseteq N(x_i)$. If $N(x_p)\subseteq N(x_i)$, then $ \{y_1,y_2, \ldots,y_j,\ldots, y_{m-1}\} \subseteq N(x_p)\subseteq N(x_i)$, which implies $y_j \in N(x_i)$. This implies that $x_iy_j\in E$, which is a contradiction. Also if $N(x_p)\supseteq N(x_i)$, then $\{y_1,y_2, \ldots ,  y_{j-1},y_{j+1},\ldots, y_{m-1},y_q\}\subseteq N(x_i)\subseteq N(x_p)$, which implies $x_py_q\in E$. Therefore, $G[X' \cup Y' \cup \{x_p,y_q\}]$ induces an $K_{m,m}-\{e\}$. Hence, if $Tr(G)=m+1$, then $G=(X\cup Y,E)$ contains either $K_{m,m}$ or  $K_{m,m}-\{e\}$ as an induced subgraph.			
		\qed
		
		From the above claim, it follows that $m>t$ contradicts the maximality of $t$. Hence $Tr(G)=t+1$.
	\end{proof}

	Next we present an algorithm to find the maximum integer $t$, such that a bipartite chain graph $G$ contains either $K_{t,t}$ or  $K_{t,t}-\{e\}$ as an induced subgraph. We show that if $t$ is the maximum index such that $G$ contains either $K_{t,t}$ or  $K_{t,t}-\{e\}$ as an induced subgraph, then first $t$ vertices of the chain ordering from each partite set induces $K_{t,t}$ or  $K_{t,t}-\{e\}$.
	
	
	\begin{lem}\label{chain_th2}
		Let $G=(X\cup Y,E)$ be a bipartite chain graph and $t$ be the maximum integer such that $G$ contains either $K_{t,t}$ or  $K_{t,t}-\{e\}$ as an induced subgraph, then $G[X_t\cup Y_t]=K_{t,t}$ or $G[X_t\cup Y_t]=K_{t,t}-\{e\}$, where $X_t=\{x_1,x_2, \ldots ,x_t\}$ and $Y_t=\{y_1,y_2,\ldots ,y_t\}$ for all $t\leq \min\{m,n\}$.
	\end{lem}
	
	\begin{proof}
		First let us assume that $t$ is the maximum integer such that $G$ contains a $K_{t,t}$, i.e., there exist $X'\subseteq X$ and $Y'\subseteq Y$ such that $G[X'\cup Y']=K_{t,t}$. If $X'=X_t$ and $Y'=Y_t$, then we are done. Otherwise, one of the following cases are true:
		
		\begin{ca}
			$X'\not= X_t$ and $Y'\not=Y_t$
		\end{ca}
		
		Since $X'\not=X_t$, there exist $p$ and $q$ with $p<q$ such that $x_p\notin X', x_q\in X'$ and $x_p\in X_t, x_q\notin X_t$ and since $Y'\not=Y_t$, there exist $i$ and $j$ with $i<j$ such that $y_i\notin Y', y_j\in Y'$ and $y_i\in Y_t, y_j\notin Y_t$. Since $i<j$, we have $N(y_i)\supseteq  N(y_j)$ and since $y_j\in Y'$ and  $G[X'\cup Y']$ is a complete bipartite graph, we have $N(y_j) \supseteq X'$. Therefore, $N(y_i)\supseteq X'$. Similarly, we have $N(x_p)\supseteq Y'$. Moreover, since $y_i\in N(x_q)$ and $N(x_q) \subseteq N(x_p)$, we have $x_py_i\in E$. Therefore, $G[X' \cup Y' \cup \{x_p,y_i\}]$ induces a $K_{t+1,t+1}$ which contradicts the maximality of $t$. 
		
		\begin{ca}
			$X'=X_t$ and $Y'\not=Y_t$
		\end{ca}
		Since $Y'\not=Y_t$, with similar arguments as above, we have $N(y_i)\supseteq X'$. This implies that $G[X_t \cup Y''] = K_{t,t}$, where $Y''=(Y'-\{y_j\})\cup \{y_i\}$. Note that $Y''$ has one more vertex common in $Y_t$ than $Y'$. So repeating this process, we finally have  $G[X_t\cup Y_t]=K_{t,t}$.
		
		\begin{ca}
			$X'\not=X_t$ and $Y=Y_t$
		\end{ca}
		Since $X'\not=X_t$, with similar arguments, we have $N(x_p)\supseteq Y'$. This implies that $G[X'' \cup Y_t] = K_{t,t}$, where $X''=(X'-\{x_q\})\cup \{x_p\}$. Note that $X''$ has one more vertex common in $X_t$ than $X'$. So repeating this process, we finally have  $G[X_t\cup Y_t]=K_{t,t}$.
		
		Hence, if $t$ is the maximum integer such that $G$ contains a $K_{t,t}$, then $G[X_t\cup Y_t]=K_{t,t}$.
		
		Next, we assume that $t$ is the maximum integer such that $G$ contains a $K_{t,t}- \{e\}$ but not a $K_{t,t}$. This implies that $t-1$ is the maximum integer such that $G$ contains $K_{t-1,t-1}$. So, by the previous result, $G[X_{t-1},Y_{t-1}]=K_{t-1,t-1}$. We show that $G[X_t\cup Y_t]= K_{t,t}-\{e\}$. Note that $N(x_t)$ contains all the vertices from the set $ \{y_1,y_2, \ldots, y_{t-1}\}$. This is because if there exists $y_i \in \{y_1,y_2, \ldots, y_{t-1}\}$ such that $y_i\notin N(x_t)$, then $y_i\notin N(x_j)$ for all $j>t$. This implies that $\{x_t,x_{t+1}, \ldots, x_m\} \notin N(y_i)$. Therefore $\{x_t,x_{t+1}, \ldots, x_m\} \notin N(y_j)$, for all $j>i$. Hence, $G$ cannot contain $K_{t,t}-\{e\}$. Similarly, $N(y_t)$ contains all the vertices from the set $ \{x_1,x_2, \ldots, x_{t-1}\}$. Also, note that $x_ty_t\notin E$ because otherwise $G[X_t \cup Y_t]= K_{t,t}$, which is a contradiction. Hence, in this case, $G[X_t \cup Y_t]= K_{t,t}-\{e\}$.
		
		Hence, if $t$ is the maximum integer such that $G$ contains either $K_{t,t}$ or  $K_{t,t}-\{e\}$ as an induced subgraph, then $G[X_t\cup Y_t]=K_{t,t}$ or $G[X_t\cup Y_t]=K_{t,t}-\{e\}$.
	\end{proof}

	The following algorithm finds the maximum integer $t$ such that $G$ contains either $K_{t,t}$ or $K_{t,t}-\{e\}$ as induced subgraph.
	
	\begin{algorithm}[h]
		
		\caption{MaxIndex($G$)}
		
		\begin{algorithmic}[1]
			
			\State  \textbf{Input:} A bipartite chain graph $G=(X\cup Y,E)$ with $X=\{x_1,x_2,\ldots ,x_{m}\}$ and $Y=\{y_1,y_2, \ldots ,y_{n}\}$ such that $N(x_m)\subseteq N(x_{m-1})\subseteq \ldots \subseteq N(x_2)\subseteq N(x_1)$ and $N(y_n)\subseteq N(y_{n-1})\subseteq \ldots \subseteq N(y_2)\subseteq N(y_1)$.
			
			\State  \textbf{Output:} Maximum $t$ such that $G$ contains either a $K_{t,t}$ or a $K_{t,t}-\{e\}$ as an induced subgraph.

			\State $ i \gets 1$, $t \gets 0$
			
			\While { $x_iy_i\in E$}
			
			\State $i=i+1$
			
			\EndWhile
			
			\State $j \gets i$
			
			\If {$x_jy_{j-1}\in E$ \& $x_{j-1}y_j\in E$ }
			
			\State $t \gets j$~~~~~~~~~~~~~~~~~~~~~ [$G$ contains $K_{t,t}-\{e\}$]
			
			\Else 
			
			\State $t\gets j-1$~~~~~~~~~~~~~~~~~[$G$ contains $K_{t,t}$]
			
			\EndIf
			
			\State \Return($t$)

		\end{algorithmic}
		
	\end{algorithm}

	Since the condition in line 8 can be checked in constant time, the above algorithm runs linearly. Hence we have the following theorem.

	\begin{theo}
		The transitivity of a bipartite chain graph can be computed in linear time.
	\end{theo}


\section{Characterization of graphs with $Tr(G)\geq t$}

In \cite{hedetniemi2018transitivity}, the authors showed that the transitivity of a graph $G$ is greater or equal to $3$ if and only if $G$ contains either $K_3$ or an induced $P_4$ or an induced $C_4$. They also posed the following open questions:

\textbf{Question 1.}
What is a necessary and sufficient condition for a graph $G$ to have $Tr(G)\geq 4$?

\textbf{Question 2.}
What is a necessary and sufficient condition for a graph $G$ to have $Tr(G)= 3$?

\noindent In this section, we present a necessary and sufficient condition for a graph $G$ to have $Tr(G)\geq t$, for any integer $t$. As a consequence of this result, we get a necessary and sufficient condition for a graph $G$ to have $Tr(G)= 3$. 

Our characterization of graphs with $Tr(G)\geq t$ is based on the result of M. Zakar, which characterizes the graphs with a Grundy number greater equal to $t$. Zakar introduced the notion of a \emph{$t$-atom} and proved that $\Gamma(G)\geq t$ if and only if $G$ contains (with respect to the \emph{cannonical partition}) a $t$-atom \cite{zaker2006results}. In transitivity, we show that the characterization can be done with subgraph containment relation. For the sake of completeness, we state the definition of \emph{$t$-atom} in this article.

\begin{defi}[\cite{zaker2006results} ]
	A $t$-atom is defined in a recursive way as follows:
	\begin{itemize}
		\item The only $1$-atom is $K_1$.
		\item The only $2$-atom is $K_2$.
		\item Let $H=(V,E)$ be any $(t-1)$-atom with $n$ vertices. Consider an independent set $\overline{K_r}$ on $r$ vertices for any $r\in \{1,2,\ldots n\}$. For that fixed $r$, consider a $r$ vertex subset $W$ of $V$ and draw a perfect matching between the vertices of $\overline{K_r}$  and $W$. Then join an edge between each vertex of $V\setminus W$ and an (and to only one) arbitrary vertex of $\overline{K_r}$. The resultant graph $G$ is a $t$-atom.
	\end{itemize}
\end{defi}


Let $\mathcal{A}_t$ denote the class of $t$-atoms. The following figure illustrates the construction of all $3$-atom.

\begin{figure}[htbp!]
	\centering
	\includegraphics[scale=0.85]{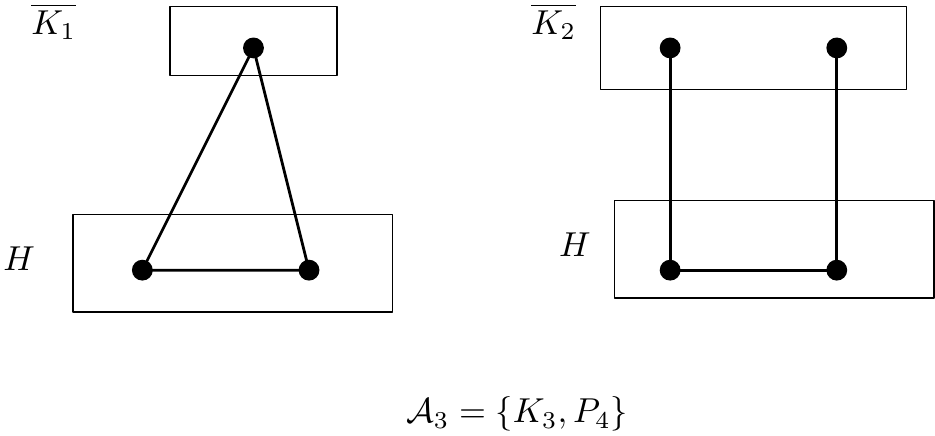}
	\caption{Construction of all $3$-atoms from $2$-atom}
	\label{Fig:atoms}
\end{figure}


Next, we prove the main theorem of this section, which characterizes the graphs with transitivity greater or equal to $t$for any integer $t$.

\begin{theo}\label{theo:GEQ4}
	\label{theo}
	For an integer $t$, the transitivity of a graph $G$ is greater or equal to $t$ if and only if $G$ contains a $t$-atom as a subgraph.
\end{theo}
\begin{proof}
	Suppose $G$ contains a $t$-atom, say $H$, as a subgraph. We show, by induction that the transitivity of $H$ is at least $t$. Clearly, transitivity of $K_1$ and $K_2$ are $1$ and $2$, respectively. Let the $t$-atom be constructed from a $(t-1)$-atom, say $H'$, and an independent set $\overline{K}_r$. By induction hypothesis, $Tr(H')=t-1$. Let us assume that $\{U_1, U_2, \ldots, U_{t-1}\}$ is a transitive $(t-1)$-partition of $H'$. By the definition of $t$-atom, $\overline{K}_r$ is a dominating set of $H'$. Therefore, $\{\overline{K}_r, U_1, U_2, \ldots, U_{t-1}\}$ forms a transitive $t$-partition of $H$. Hence,  the transitivity of $H$ is at least $t$. Since $H$ is a subgraph of $G$, then $Tr(G)\geq Tr(H)\geq t$.
	
	Conversely, let us assume that the transitivity of $G=(V,E)$ is greater or equal to $t$. Therefore, $G$ has a transitive $t$-partition \cite{hedetniemi2018transitivity}.  Let $\{V_1,V_2,\ldots,V_t\}$ be a transitive $t$-partition of $G$. Once again , by induction, we show that $G$ contains a $t$-atom. For the base case $t=1$, the statement is trivially true. Note that transitivity of $G'\geq t-1$, where $G'=G\setminus \{V_1\}$. By induction hypothesis, $G'$ contains a $(t-1)$-atom, say $H=(V_H, E_H)$. Now, $V_1$ is a dominating set of $G$, therefore $V_1$ dominates every vertex of $V_H$. Let us consider a subset of vertices $B\subseteq V_1$ such that $B$ dominates every vertex of $V_H$ and $\lvert B \rvert$ $\leq$ $\lvert V_H \rvert$. By Hall's theorem \cite{diestel2005graph}, there exists a matching, say $M$, between $B$ and $V_H$ of size $|B|$. Let $W$ be the endpoints of $M$ that are in $V_H$. Since, $B$ dominates every vertex of $V_H$, every vertex of $V_H\setminus W$ has a neighbour in $B$. Let $X$ be the set of edges between $B$ and $V_H\setminus W$ such that every edge of $X$ is incident to exactly one vertex in $V_H\setminus W$. Now, a $t$-atom can be obtained from $G$ by removing the vertices $(V\setminus \{B \cup V_H\})$ and removing all the edges $(E \setminus \{E_H \cup M\cup X\})$.
\end{proof}


\begin{remk}\label{rem:UsingSubgraph}
	\textnormal{Note that the class of graphs that contains $K_3$ or $P_4$ as a subgraph is equivalent to the class of graphs that contains $K_3$ or $P_4$ or $C_4$ as an induced subgraph. Therefore, the characterization of graphs with transitivity greater or equal to $3$, proved in \cite{hedetniemi2018transitivity}, is a special case of Theorem \ref{theo:GEQ4}.}
\end{remk}

\begin{figure}[htbp!]
	\centering
	\includegraphics[scale=0.65]{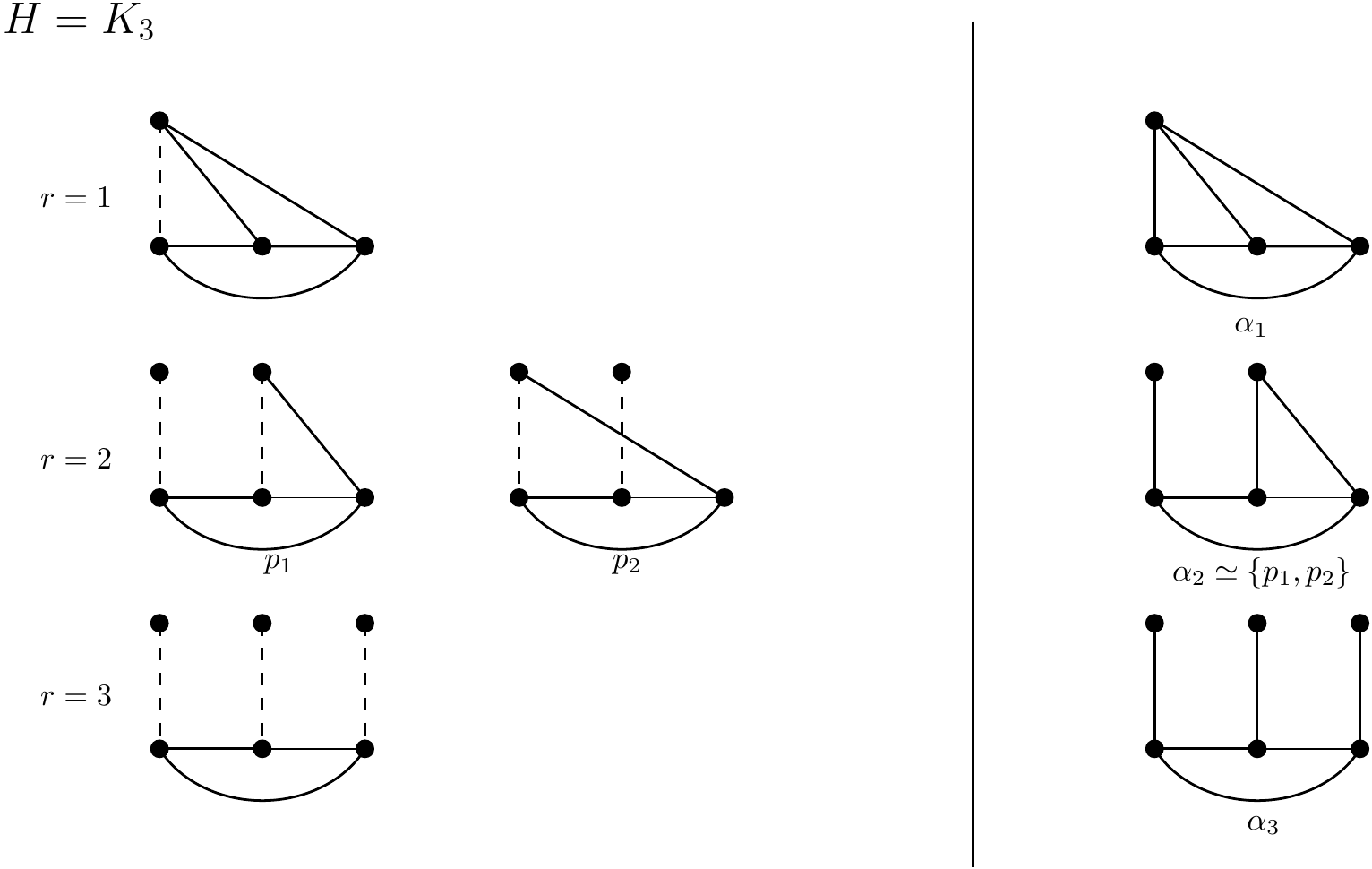}
	\caption{Non-isomorphic $4$-atoms obtained from $K_3$}
	\label{fig:K_3subgraphs}
	
\end{figure}
\begin{figure}[htbp!]
	\centering
	\includegraphics[scale=0.65]{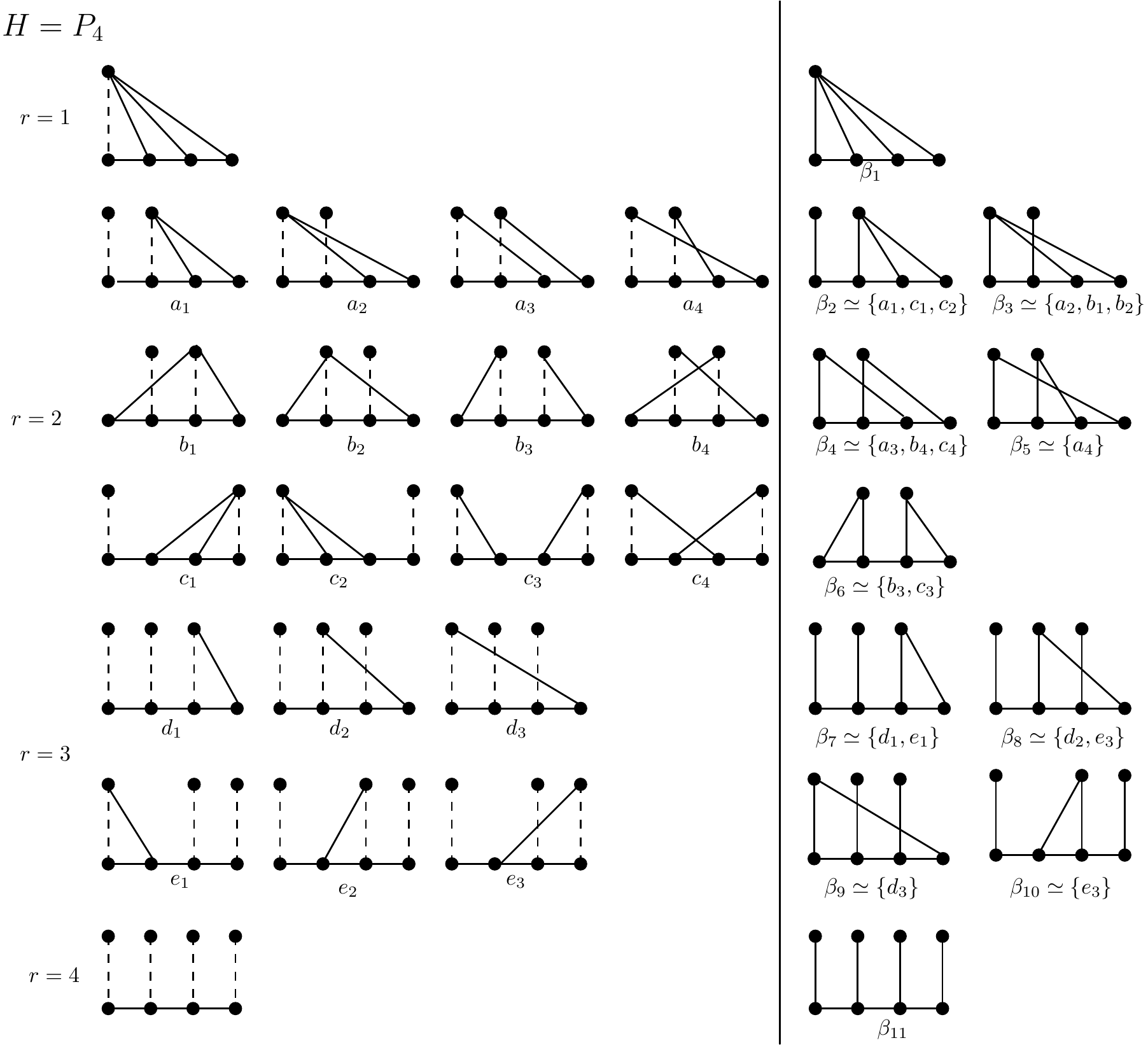}
	\caption{Non-isomorphic $4$-atoms obtained from $P_4$}
	\label{fig:P_4subgraphs}
\end{figure}

Next we list all non-isomorphic copies of $\mathcal{A}_4$. Figure \ref{fig:K_3subgraphs} and Figure \ref{fig:P_4subgraphs} illustrates the $4$-atoms that can be obtained from the two $3$-atoms, namely $K_3$ and $P_4$, respectively.

Note that the graph $\alpha_2$ is a subgraph of $\beta_2$. This implies that the class of graphs containing a member of $\mathcal{A}_4$ is the same as the class of graphs containing a member from $\mathcal{A}$, where $\mathcal{A}= \mathcal{A}_4 \setminus \{\beta_2\}$. Hence, we have the following corollary, which answers \emph{Question 1}.

\begin{coro}\label{coro:geq4}
	The transitivity of a graph $G$ is greater or equal to $4$ if and only if $G$ contains one of the graph form $\mathcal{A}$ as a subgraph.
\end{coro}


Since, by Remark \ref{rem:UsingSubgraph}, the transitivity of a graph $G$ is greater or equal to $3$ if and only if $G$ contains either $K_3$ or $P_4$ as subgraph, we have the following corollary which answers \emph{Question 2}.

\begin{coro}\label{coro:eq3}
	The transitivity of a graph $G$ is equal to $3$ if and only if $G$ contains either $K_3$ or $P_4$ as subgraph but does not contain any graph from $\mathcal{A}$ as a subgraph.
\end{coro}

\begin{remk}
	Since, given a graph $G$, in polynomial time, we can check whether another fixed graph $H$ is a subgraph of $G$ or not, the conditions in Corollary \ref{coro:geq4} and \ref{coro:eq3} can also be done in polynomial time.
\end{remk}


\section{Conclusion}
In this paper, we have shown that \textsc{Maximum Transitivity Decision Problem} is NP-complete for bipartite graphs. On the positive side, we have demonstrated that the transitivity of a given bipartite chain graph can be computed in linear time. Then, we have provided a necessary and sufficient condition for a graph to have transitivity greater or equal to $t$for any integer $t$. It would be interesting to investigate the complexity status of this problem in other subclasses of bipartite graphs. Designing an approximation algorithm for this problem would be another challenging open problem.

\bibliographystyle{alpha}
\addcontentsline{toc}{section}{Bibliography}
\bibliography{Transitivity_reference}

\end{document}